\documentstyle[prc,aps,preprint]{revtex}
\title{{\small {\bf The $K^{+}\rightarrow \pi^{+}\nu\bar{\nu}$ Rare Decay in 
Two Higgs Doublet Model }}}
\author{T. BARAKAT\\
{\it {\small Civil Engineering Department, Near East University }}\\
{\it {\small Lefko\c{s}a, Mersin- Turkey }}}
\date{}
\begin{document}
\begin{titlepage}
\maketitle
\begin{abstract}
\baselineskip .9cm
The rare $K^{+}\rightarrow \pi^{+}\nu\bar{\nu}$ decay is investigated in the 
context of type II two-Higgs-doublet model (2HDM). By using the existing 
experimental data of the branching ratio, restrictions on the free parameters 
of the model $m_{H}$, and $tan\beta$ are obtained: 
$0.7\leq tan\beta \leq 0.8$, and 
$500 GeV\leq m_{H} \leq 700GeV$. 
\end{abstract}
\end{titlepage}
\textwidth 16cm
\rightmargin 4.cm
\parskip .5cm
\baselineskip .9cm
\section{ Introduction}
\hspace{0.6cm} The determination of the elements of the Cabibbo-Kobayashi
-Maskawa matrix (CKM) is still an important issue in the flavor 
physics. The precise determination of the CKM parameters will be one of the 
most important progresses to understand the nature, physics of violated 
symmetry.  

In this sense the rare   $K^{+}\rightarrow \pi^{+}\nu\bar{\nu}$ decay has 
attached a special interest due its sensitivity for the determination of 
CKM parameters, in particular the element $V_{td}$, and considered one 
of the cleanest decays from a theoretical standpoint. Moreover this decay 
occupies a special place, since for this decay the short distance effects 
dominated over the long distance effects. Over the years important 
refinements have been added in the theoretical treatment of 
$K^{+}\rightarrow \pi^{+}\nu\bar{\nu}$,  long-distance contributions to the 
branching ratio were estimated quantitatively and could be shown to be 
essentially negligible as expected, two to three orders of magnitude smaller 
than the short distance contribution at the level of the branching ratio [1]. 
On the other hand, the calculation [2] of next-to-leading QCD correction 
reduced considerably the theoretical uncertainty due to the choice of the 
renormalization scales present in the leading order expression. Since the 
relevant hadronic matrix element of the operator 
$\bar{s}\gamma_{\mu}(1-\gamma_{5})d\bar{\nu}\gamma_{\mu}(1-\gamma_{5})\nu$ 
can be extracted in the leading decay 
$K^{+}\rightarrow \pi^{0} e^{+}\nu$. Conventionally, the 
Br($K^{+}\rightarrow \pi^{+}\nu\bar{\nu}$) is related to the experimental 
well-known quantity Br($K^{+}\rightarrow \pi^{+}e^{+}\nu$)=0.0482, measured 
to ${1\%}$ accuracy. The resulting 
theoretical expression for Br($K^{+}\rightarrow \pi^{+}\nu\bar{\nu}$) is only 
a function of the CKM parameters, the QCD scale $\Lambda_{\bar{M}s}$ and 
the quark masses $m_{t}$ and $m_{c}$. 

Experiments in the K meson system have entered new period. That the branching 
ratio of the flavor-changing neutral current (FCNC) 
process $K^{+}\rightarrow \pi^{+}\nu\bar{\nu}$ has been recently measured, 
and it has turned out to be 
$Br(K^{+}\rightarrow \pi^{+}\nu\bar{\nu})=(4.2^{+9.7}_{-3.5}).10^{-10}$ [3]. 
The central value seems to be 4-6 times larger than the predictions of the 
Standard Model (SM) Br(0.6-1.5)x$10^{-10}$ [4]. Hence the rare  
$K^{+}\rightarrow \pi^{+}\nu\bar{\nu}$ decay is very sensitive 
to a new physics beyond the SM. Therefore, careful investigation 
of this decay can provide useful information about new physics [5]. For this 
reason different new physics scenarios for this decay will become very actual.
 
In this work the decay $K^{+}\rightarrow \pi^{+}\nu\bar{\nu}$ is investigated 
in the framework of the two Higgs doublet model (2HDM). We estimate the 
constraints of the 2HDM parameters namely, $tan\beta$ and $m_{H}$, using 
the result coming from the measurement of [3]. Subsequently, this paper is 
organized as follows: In Section 2, the relevant effective 
Hamiltonian for the decay $K^{+}\rightarrow \pi^{+}\nu\bar{\nu}$ in 2HDM is 
presented. Section 3, being devoted to the numerical analysis of our 
results; and finally a brief discussion of the results is given. 
\section{Effective Hamiltonian}
\hspace{0.6cm} In the Standared Model  (SM) the process 
$K^{+}\rightarrow \pi^{+}\nu\bar{\nu}$ is described at quark level by the 
$s\rightarrow d\nu\bar{\nu}$ transitions and received contributions from 
$Z^{0}$-penguin and box diagrams. The effective Hamiltonian relevant to  
$s\rightarrow d\nu\bar{\nu}$ transition is described by only one Wilson 
coefficient, and its explicit form is:   
\begin{eqnarray}
H_{eff}&=&\frac{G}{\sqrt{2}}\frac{ \alpha}{2\pi sin^{2}\theta_{w}}V^{*}_{ts}
V_{td}C_{11}^{SM}\bar{s}\gamma_{\mu}(1-\gamma_{5})d\bar{\nu}\gamma_{\mu}
(1-\gamma_{5})\nu,
\end{eqnarray}
where G is the Fermi coupling constant, $\alpha$ is the fine structure 
constant, $V_{tb}V^{*}_{ts}$ are products of Cabibbo-Kabayashi-Maskawa 
matrix elements and $x_{t}=\frac{m_{t}^{2}}{m_{w}^{2}}$. The resulting 
expression of Wilson coefficient $C_{11}$, which
was derived in the context of the SM including $O(\alpha_{s})$ corrections is 
[6,7]  
\begin{eqnarray}
C_{11}^{SM}=\left[X_{0}(x)+\frac{\alpha_{s}}{4\pi}X_{1}(x)\right],
\end{eqnarray} 
with 
\begin{eqnarray}
X_{0}(x)=\eta\frac{x}{8}\left[\frac{x +2}{x-1}+\frac{3x-6}
{(x-1)^{2}}lnx\right],
\end{eqnarray}    
\begin{eqnarray}
X_{1}(x)&=&\frac{4x^{3}-5x^{2}-23x}{3(x-1)^{2}}-
\frac{x^{4}+x^{3}-11x^{2}+x}{(x-1)^{3}}lnx+
\frac{x^{4}-x^{3}-4x^{2}-8x}{2(x-1)^{3}}ln^{2}x \nonumber \\
&+&\frac{x^{3}-4x}{(x-1)^{2}}Li_{2}(1-x)+8x\frac{\partial X_{0}(x)}{\partial x}
lnx_{\mu}.
\end{eqnarray}
Here $Li_{2}(1-x)=\int_{1}^{x}\frac{lnt}{1-t}dt$ and 
$x_{\mu}=\frac{\mu^{2}}{m_{w}^{2}}$ with $\mu=O(m_{t})$.

At $\mu=m_{t}$, the QCD correction for $X_{1}(x)$ term is very small (around 
$\sim 3\%$), and $\eta =0.985$ is the next-to-leading order (NLO) QCD 
correction to the t- exchange calculated in [2]. 

From the theoretical point of view, the transition 
$s\rightarrow d\nu\bar{\nu}$ is a very clean process as pointed out, since it 
is practically free from the scale dependence, and free from any long 
distance effects. In addition, the presence of a single operator 
governing the inclusive $s \rightarrow d \nu\bar{\nu}$ transition is an 
appealing property. The theoretical uncertainty within the SM is only related 
to the value of the Wilson coefficient $C_{11}$ due to the 
uncertainty in the top quark mass. In this work, we have considered possible 
new physics in $s \rightarrow d \nu\bar{\nu}$ only through the value of that 
Wilson coefficient.
 
 In this spirit, the process $s\rightarrow d\nu\bar{\nu}$ in the context of 
the 2HDM  has additional contributions from $Z^{0}$-penguin and box diagrams 
through H boson exchanges. The relevant Feynman diagrams correspond to the 
transition $s\rightarrow d\nu\bar{\nu}$ has been given in [8,9]. The first 
three diagrams describe the effective Hamiltonian in the SM, while the last 
three diagrams represent the 2HDM contributions 
to the $s\rightarrow d\nu\bar{\nu}$ transition, due to the charged Higgs 
boson exchanges. The interaction lagrangian between the 
charged Higgs bosons fields and fermions are then given by:
\begin{eqnarray}
  L=(2\sqrt{2}G_{F})^{1/2}\left[ tan\beta\bar{U_{L}} V_{CKM}M_{D}D_{R}+
ctg\beta\bar{U_{R}}M_{U}V_{CKM}D_{L}+tan\beta\bar{N_{L}}M_{E}E_{R}\right ] 
H^{+}+h.c.
\end{eqnarray}
Here, $H^{+}$ represents the charged physical Higgs field. $U_{L}$ and $D_{R}$ 
represent left-handed up and right-handed down quark fields. $N_{L}$ and 
$E_{R}$  are left-handed neutral and right-handed charged leptons.
$M_{D}$, $M_{U}$, and $M_{E}$ are the mass matrices for the down quarks, up 
quarks, and charged leptons respectively. $V_{CKM}$ is the Cabibbo-Kobayashi-
Maskawa matrix. $tan\beta$-is the ratio of the vacuum expectation values of 
the two Higgs doublets in 2HDM, and it is a free parameter of the model. 

 From eq.(5), it follows that the box diagrams contribution to the process 
$s\rightarrow d\nu\bar{\nu}$ in 2HDM are proportional to the charged lepton 
masses; and therefore, they are giving a negligible contribution.
So in this model, the transition $s\rightarrow d\nu\bar{\nu}$ in eq.(1) can 
only include extra contribution due to the charged Higgs interactions. Hence, 
the charged Higgs contribution modify only the value of the Wilson 
coefficient $C_{11}$ (see eq.(1)), and it does not induce any new operators 
(see also [8,9]):
\begin{eqnarray}
C_{11}^{ 2HDM}&=&-\frac{1}{8}xy ctg^{2}\beta \left\{\frac{1}{y-1}-
\frac{lny}{(y-1)^{2}}\right\},
\end{eqnarray}
where $x=\frac{m^{2}_{t}}{m^{2}_{W}}$  and  $y=\frac{m^{2}_{t}}{m^{2}_{H}}$.

As we noted earlier the QCD corrections practically do not change the value of 
$C_{11}$. If so, eq. (2) and eq.(6), are plugged in eq.(1), to obtain a 
modified effective Hamiltonian, which represents 
$s \rightarrow d \nu\bar{\nu}$ decay in 2HDM: 
\begin{eqnarray}
H_{eff}=\frac{G}{\sqrt{2}}\frac{\alpha}{2\pi sin^{2}\theta_{w}}V_{td}V^{*}_{ts}
[X_{tot}]\bar{s}\gamma_{\mu}(1-\gamma_{5})d\bar{\nu}\gamma_{\mu}
(1-\gamma_{5})\nu,  
\end{eqnarray}
where $X_{tot}= C_{11}^{SM}+C_{11}^{2HDM}$.

However, in spite of such theoretical advantages, it would be a very difficult 
task to detect the inclusive $s \rightarrow d \nu\bar{\nu}$  decay 
experimentally, because the final state contains two missing neutrinos and 
many hadrons. Therefore, only the exclusive channels are expected, namely 
$K^{+} \rightarrow \pi^{+} \nu\bar{\nu}$, are well suited to search for and 
constrain for possible "new physics" effects.  

In order to compute $K^{+} \rightarrow \pi^{+}\nu\bar{\nu}$ decay, we need the 
matrix elements of the effective Hamiltonian eq.(7) between the final and 
initial meson states. This problem is related to the non-perturbative sector 
of QCD and can be solved only by using non-perturbative methods. The matrix 
element  $<\pi^{+} \mid H_{eff}\mid K^{+}>$ has been investigated in a 
framework of different approaches, such as chiral perturbation theory [10], 
three point QCD sum rules [11], relativistic quark model by the light front 
formalism [12], effective heavy quark theory [13], and light cone QCD sum 
rules [14,15]. As a result, the hadronic matrix element for the 
$K^{+} \rightarrow \pi^{+} \nu\bar{\nu}$ can be parameterized in terms of form 
factors:
\begin{eqnarray}
<\pi \mid \bar{s}\gamma_{\mu}(1-\gamma_{5})d\mid K>
&=& f_{+}^{\pi^{+}}(q^{2})(p_{K}+p_{\pi})_{\mu}+f_{-}q_{\mu},
\end{eqnarray}   
where  $q_{\mu}=p_{K}-p_{\pi}$, is the momentum transfer. In our calculations
the form factor $f_{-}$ part do not give any contributions since its 
contribution $\sim m_{\nu}$=0.
After performing the mathematics and taking into 
account the number of light neutrinos $N_{\nu}=3$ the differential 
decay width is expressed as:  
\begin{eqnarray}
\frac{d\Gamma(K^{+} \rightarrow \pi^{+} \nu\bar{\nu})}{dq}&=&
\frac{G^{2}\alpha^{2}\eta^{2}}
{2^{8}\pi^{5}sin^{4}\theta_{w}}m_{K^{+}}^{3}\mid V_{tb}V^{*}_{ts}\mid^{2}
\lambda^{3/2}(1,r_{+},s)
\mid X_{tot} \mid^{2} \mid f_{+}^{\pi^{+}}(q^{2})\mid^{2},
\end{eqnarray}
where  $r_{+}=\frac{m^{2}_{\pi^{+}}}{m^{2}_{K^{+}}}$  and   
$s=\frac{q^{2}}{m^{2}_{K^{+}}}$.

Similar calculations for  $K^{+}\rightarrow \pi^{0} e^{0}\bar{\nu}$ 
lead to the following result:
\begin{eqnarray}
\frac{d\Gamma(K^{+} \rightarrow \pi^{0} e^{+}\bar{\nu})}{dq}&=&
\frac{G^{2}}{192\pi^{3}}\mid V_{us}\mid^{2} \lambda^{3/2}(1,r_{-},s)
m_{K^{+}}^{3}\mid f_{+}^{\pi^{0}}(q^{2})\mid^{2},
\end{eqnarray}
where  $r_{-}=\frac{m^{2}_{\pi^{0}}}{m^{2}_{K^{+}}}$, and 
$\lambda (1,r_{\pm},s)=1+r_{\pm}^{2}+s^{2}-2r_{\pm}s-2r_{\pm}-2s$ is the usual 
triangle function.
 In derivation of eq.(10) we neglect the electron mass, and we use the form 
factors $f_{+}^{\pi^{+}}=\sqrt{2}f_{+}^{\pi^{0}}$ which follows from isotopic 
symmetry. Using eq.(9) and eq.(10)
 one can relate the branching ratio of  $K^{+}\rightarrow \pi^{+}\nu\bar{\nu}$ 
to the well known measured decay $K^{+} \rightarrow \pi^{0} e^{0}\bar{\nu}$
branching ratio:
\begin{eqnarray}
B(B^{+}\rightarrow K^{+}\nu\bar{\nu})=k \left[\left( \frac{Im \lambda_{t}}
{\lambda^{5}}X_{tot} \right)^{2}+\left(\frac{Re \lambda_{c}}{\lambda}P_{0}
(K^{+})+\frac{Re \lambda_{t}}{\lambda^{5}}X_{tot}\right)^{2}\right],
\end{eqnarray}
where 
$k=r_{K^{+}}\frac{3\alpha^{2}B(K^{+} \rightarrow \pi^{0} e^{+}\nu)}{2\pi^{2} 
sin^{4}\theta_{w}}\lambda^{8}=4.11.~10^{-11}$.

Here $r_{K^{+}}=0.901$ summaries the isospin-breaking corrections which 
come from phase space factors due to the difference of masses of $\pi^{+}$ and 
$\pi^{0}$.

In derivation eq.(11) we have used the wolfenstein parametrization of the CKM 
matrix, in which each element is expanded as a power series in the small 
parameter \\
$\lambda =\mid V_{us}\mid$=0.22, $\lambda_{i}=V^{*}_{is}V_{id}$ and 
$P(K^{+})$ represent the sum of charm contributions to the two diagrams 
including the (NLO) QCD corrections [2]. At $m_{c}=1.3$ GeV, 
$\Lambda_{\bar{M}s}=0.325 $ GeV and at renormalization scale $\mu_{c}=m_{c}$ 
in [16] it is found that $P_{0}(K^{+})=0.4 \pm 0.06$. 

\section{Numerical Analysis}
In the numerical analysis, the following values have been used as input 
parameters:\\
$G_{F}=1.17{~}.10^{-5}~ GeV^{-2}$, $\alpha =1/137$, and $\lambda=0.22$
As we noted early we used the wolfeustein parametrization of CKM matrix 
elements. In this parametrization \\
$Im \lambda_{t}=A^{2}\lambda^{5}\eta$,~~~~
$Re \lambda_{c}=-\lambda (1-\lambda^{2}/2)$, and
$Re \lambda_{t}=-A^{2}\lambda^{5}(1-\rho)$.

The parameter A determines from $b\rightarrow c$ transition and its 
$ A=0.80 \pm 0.075$  [17]. The other two CKM parameters $\rho$ and $\eta$ 
are constrained by the measurements of $\mid V_{ub}/V_{cb} \mid$, 
$x_{d}(B^{0}_{d}- \bar{B^{0}_{d}})$  mixing, and  $\mid \epsilon \mid$ (the CP 
violation parameter in the kaon system). For typical values of the necessary 
input parameters of   $\rho$ and $\eta$ we have adopt the following two sets 
:\\
\begin{eqnarray}
\mbox{set I}: \left\{\begin{array}{l}    
             \rho =0.06 \\              
             \eta =0.35         
       \end{array}
            \right.
~~~~~~~~~~~~~~~~\mbox{set II}: \left\{\begin{array}{l}
             \rho =-0.25 \\
             \eta =0.3
            \end{array}
              \right..
\end{eqnarray}

The free parameters of the 2HDM model which we have used namely 
$tan\beta$ and $m_{H}$ are not arbitrary, but there are some constraints on 
them by using the existing experimental data. These constraints are usually 
obtained from $B^{0}-\bar{B}^{0}$, 
$K^{0}-\bar{K}^{0}$ mixings, $b\rightarrow s\gamma$ decay width, 
$R_{b}=\frac{\Gamma(z\rightarrow b\bar{b})}{\Gamma(z\rightarrow hadrons)}$, 
and semileptonic  $b\rightarrow c\bar{\nu_{\tau}}\tau$ decay which are given 
by [18] as
\begin{equation} 
0.7\leq tan\beta \leq 0.6 (\frac{m_{H}^{+}}{1GeV}),
\end{equation}
where as a lower bound for the charged Higgs mass  $m_{H}\geq 300$ GeV at 
$\mu=5$ scale has been estimated in 2HDM [19]. If these constraints are 
respected, an upper and lower bound for $ctg\beta$ is extracted:
\begin{eqnarray}
0.004\leq ctg\beta \approx 2.
\end{eqnarray}
In Figures 1 and 2, we represent the branching ratio of 
$K^{+} \rightarrow \pi^{+} \nu\bar{\nu}$ as a function of $ctg\beta$ 
for various values of $m_{H}$, and as a function of $m_{H}$ for various values 
of $ctg\beta$. For illustrative purposes we consider three values of 
$ctg\beta$, namely $ctg\beta=$1, 1.5 and 2 and we allow $m_{H}$ to range 
between 300 GeV and 1000 GeV, and then we consider three values of $m_{H}$, 
namely $m_{H}$=300, 500, 1000 GeV and we allow $ctg\beta$ to range between 0 
to 2.  It can be seen that for $ctg\beta=1$, the branching ratio (BR) for 
$K^{+} \rightarrow \pi^{+} \nu\bar{\nu}$ decay increases slowly 
with the increasing of $m_{H}$; whereas, for larger values of $ctg\beta$, 
the BR decreases at all values of $m_{H}$. Furthermore, when the $ctg\beta$ 
is increased the BR rapidly grows up. Therefore, it can be 
concluded that the main contribution to the decay width comes from the charged 
Higgs exchange diagrams (see [8,9]).

The question now is; what kind of restrictions on $tan\beta$ and $m_{H}$
can be obtained if the recent experimental result of 
$Br(K^{+}\rightarrow \pi^{+}\nu\bar{\nu})=(4.2^{+9.7}_{-3.5}).~10^{-10}$ [3] 
is respected that is:
\begin{equation} 
(0.7\leq BR^{exp.} \leq 13.9).~10^{-10},
\end{equation}
and whether or not it coincide with the restrictions given in [18]. For this 
aim, in Figure 3 we present the dependence of $tan\beta$ on  $m_{H}$ using 
both sets of values of $\rho$ and $\eta$. We see that when 
$300 GeV \leq m_{H} \leq 1 TeV$ it gives:
\begin{equation}
0.18\leq tan\beta \leq 0.5\pm 0.2~~~~~~~~  (set I),
\end{equation}
\begin{equation}
0.18\leq tan\beta \leq 0.8\pm 0.3 ~~~~~~~~~(set II). 
\end{equation}
If we use the lowest bound for $tan\beta=0.7$ (see eq.(13)) we  see that the 
set I predictions is ruled out and for set II  we have small room for 
$tan\beta$, namely from eq.(16) and from eq.(17) we have: 
\begin{equation} 
0.7\leq tan\beta \leq 0.8.
\end{equation}
If we increase a little bit the upper bound and if we put a lower 
value for $m_{H}$=500 GeV we can see that in this case
\begin{equation} 
0.7\leq tan\beta \leq 0.9.
\end{equation}

Using these results we can conclude that the mass of the charged Higgs boson 
must be lie in the interval 
\begin{equation} 
500 GeV \leq m_{H} \leq 700 GeV.
\end{equation}

In conclusion, using the experimental result of the branching ratio for 
$K^{+}\rightarrow \pi^{+}\nu\bar{\nu}$ and the CLEO measurements on
$b\rightarrow s\gamma$ [20] we find new restrictions on the free parameters 
$tan\beta$ and $m_{H}$ of the 2HDM model. In summay it is found that the 
contribution of type II two-Higgs-doublet model to the branching ratio is 
exceed at most by $\sim 20\%$ from the standard model ones. 
\pagebreak
\begin{center}
Figure Captions
\end{center}
~~~~\\
Figure 1 : The dependence of the Br($K^{+}\rightarrow \pi^{+}\nu\bar{\nu}$) 
             on  $ctg\beta$ at fixed values of $m_{H}$.\\
Figure 2  : The dependence of the Br($K^{+}\rightarrow \pi^{+}\nu\bar{\nu}$)
             on $m_{H}$ at fixed values of $ctg\beta$.\\
Figure 3 : The dependence of $tan\beta$ on the charged Higgs boson mass 
          $m_{H}$. Curves (A, B), and (C, D) describes upper 
          and lower bound of the experimental values of the 
          Br($K^{+}\rightarrow \pi^{+}\nu\bar{\nu}$) for Set I and 
          Set II values of $\rho$ and $\eta$ respectively.
 
\pagebreak

\end{document}